# Probing Laser-induced Microenvironment Changes in Room Temperature Ionic Liquids

Renjun Ma,[1,2] Xian Wang,[1,2] Jialong Jie,[1,2] Linyin Yan,[1,3] Zhuoran Kuang,[1,2] Qianjin Guo,*[,1,2] Boxuan Li*[,1,2] and Andong Xia*[,1,2]

[1]*Beijing National Laboratory for Molecular Sciences (BNLMS), Key Laboratory of Photochemistry, Institute of Chemistry, Chinese Academy of Sciences, Beijing 100190, People's Republic of China*

[2]*University of Chinese Academy of Sciences, Beijing 100049, People's Republic of China*

[3]*Reed Elsevier Information Technology (Beijing) Co., Ltd, Beijing 100738, People's Republic of China*

Corresponding Authors: andong@iccas.ac.cn; lbx1984@iccas.ac.cn; guoqj@iccas.ac.cn

**Abstract:**

Modulating heterogeneous microstructure in room temperature ionic liquids (RTILs) by external stimuli is an important approach for understanding and designing the external field induced chemical reactions in natural and applicable systems. Here, we report for the first time the redistribution of oxygen molecules in RTILs due to laser-induced microstructure changes probed by triplet excited state dynamics of porphyrin and rotational dynamics of coumarin 153. A remarkably long-lived triplet excited state of porphyrin is observed in air-saturated ionic liquid with the changes of microstructure after irradiation, suggesting that more charge-shifted $O_2$ induced by external laser field move into the polar domains of ionic liquid [$C_8$mim][$PF_6$] from nonpolar domains through electrostatic interactions. The results presented here suggest that the heterogeneous systems of ionic liquids upon external stimuli can be designed for those oxygen-related chemical reactions with extensive inspirations for potential applications in lithium-air batteries, gaseous sensing, photoelectrical catalysis and so on.

Photocatalysis, aerobic reaction, energy/charge transfer and many natural processes are all involved with physical chemistry behaviors of solutes and/or solvents in a heterogeneous microenvironment.[1-9] Triggered by external optical or electrical control methods to modulate the physical chemistry processes in various heterogeneous systems is a high interest research subject today.[10-14] Many probe molecules are optimized to improve their responsive properties for investigating the changes of their surrounding microenvironment.[15] A heterogeneous solvent could provide a more versatile media, where simultaneous modulating the microenvironment of solvent and the associated physical chemistry properties (such as spectral behaviors) of solute would possibly shed a new light for many promising applications.

Room temperature ionic liquids (RTILs) as alternatives to conventional solvents have been extensively investigated due to their many advantages like almost non-volatile, high thermal stability, broad electrochemical window, which emerged as benign media for reaction, catalysis, electrolytes, gas adsorption and so on.[16-20] Especially, one of the most unique natures of ionic liquids is their structural heterogeneity which provides a microenvironment with potential designer heterogeneous microstructures composed of all cations and anions for photo-electric control.[21-25] Both experimental and computational evidences show that some organized structures with different structural domains could form within ionic liquids by their polar and nonpolar components.[26-31] These instinct heterogeneous microenvironment of RTILs can be sensitively revealed by the dynamics of molecular probes through many spectroscopic methods, which has been studied by many research groups.[32-52] Porphyrins with their excellent triplet excited state properties are important candidates for the photochemical and photoelectric researches related to the applications of solar cells, sensing, diagnostics and therapeutics.[52-56] Moreover, because the dynamics of their triplet excited states are very sensitive to the microenvironment associated with oxygen, porphyrin molecules are widely used as a probe to study the microstructure changes associated with oxygen-related chemical reactions in many heterogeneous systems.[57-59]

By now, how to modulate and identify the heterogeneous microstructure changes of ionic liquids for their applications is still a challenge. It is of great interest to see how this spatial heterogeneity within ionic liquids changes when an external stimuli is applied. In this study, we demonstrate a successful modulation and identification of triplet excited state dynamics of *meso*-tetraphenylporphyrin (TPP) from a structural heterogeneity change of ionic liquids with the redistribution of oxygen induced by external laser field, which is also proved by the complementary results of rotational dynamics of coumarin 153 (C153) and molecular dynamics (MD) simulations. As neutral molecules like TPP and C153, they can well disperse into the heterogeneous microenvironment of ionic liquids. By the controllable laser-induced means, the spectral behavior of these molecular probes can be used to determine the structural heterogeneity changes of ionic liquids and oxygen molecules redistribution in the heterogeneous microenvironment by external field. The triplet excited state relaxations of TPP in ionic liquids are mainly affected by the surrounding oxygen molecules, which can be a sensitive reporter for the oxygen redistribution with the changed microstructure in heterogeneous ionic liquids upon external laser irradiation. The molecular structures of ionic liquids $[C_n\text{mim}][PF_6]$ ($n$ = 4, 6, 8), and TPP and C153 molecules used in this study are shown in Scheme 1.

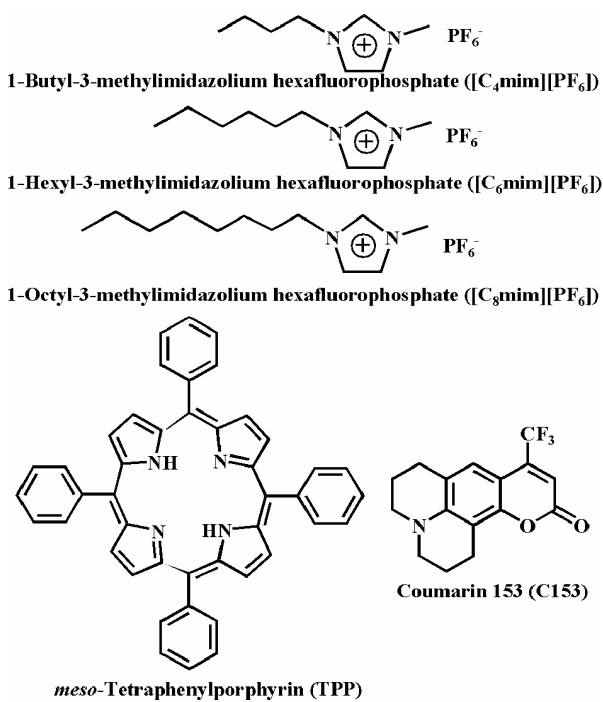

**Scheme 1 | Molecular structures of ionic liquids and molecular probes.** Ionic liquids for [$C_n$mim][$PF_6$] (*n* = 4, 6, 8), and TPP and C153 molecules.

## Results

**Spectroscopic behaviors of *meso*-tetraphenylporphyrin (TPP).** 1-Alkyl-3-methylimidazolium hexafluorophosphate ([$C_n$mim][$PF_6$]) is one of the most commonly used ionic liquids with a structural heterogeneity from the interconnected and permeated polar and nonpolar domains when the alkyl chain is butyl or longer (*n* ≥ 4). Polar domains are formed by the charge ordering of the imidazolium rings of cations and the anions through Coulomb electrostatic interactions, and nonpolar domains arise from the aggregation of alkyl chains through van der Waals interactions.[29, 30] For ionic liquid [$C_8$mim][$PF_6$] as shown in Scheme 1, the long octyl chains can come close to each other in an ordered interdigitated way or individually coil themselves remaining a relatively large free volume as investigated by our previous study.[30] All the ions of ionic liquid [$C_8$mim][$PF_6$] have the possibility to change their arrangement upon external stimuli such as light or electric fields, which could be an appropriate heterogeneous media for orderly induced modulations.[25]

As neutral molecules, TPP could possibly well disperse into the polar and nonpolar domains of ionic liquid [$C_8$mim][$PF_6$] and experience the heterogeneous structure. The absorption spectra of TPP in ionic liquid [$C_8$mim][$PF_6$] have two well-defined regions, which are B (known as Soret) bands and Q-bands as shown in Fig. 1a. The sharp single peak of B-band and the presence of the four Q-bands (inset of Fig. 1a) indicate that there is no aggregation of TPP molecules in ionic liquid.[60] Fig. 1b shows the transient absorption spectrum (at 7 μs) of triplet excited state of TPP (positive ESA peak at ~450 nm) and ground state depletion (negative GSD peak at ~417 nm) in air-saturated ionic liquid upon 532 nm excitation. With no obvious transient absorption observed from the neat ionic liquid, the influences of neat ionic liquid are negligible as shown in Fig. 1b.

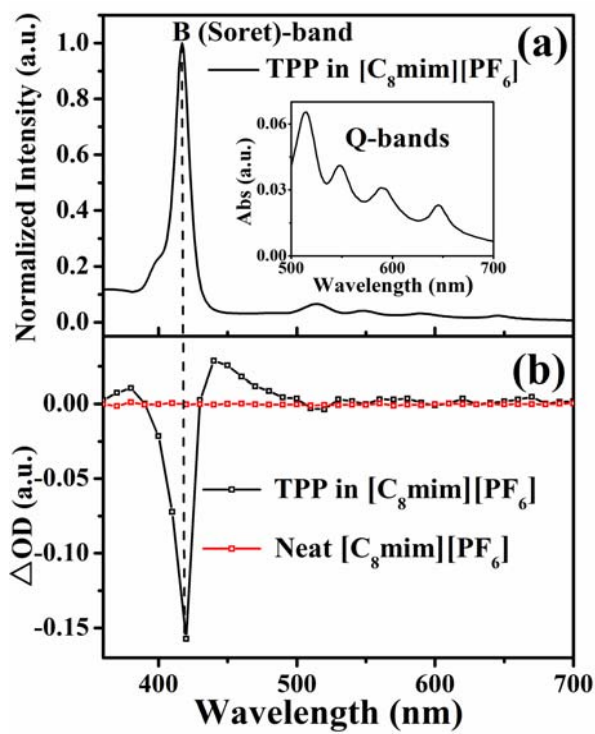

**Figure 1 | Steady-state and transient-state absorption spectra of TPP in ionic liquid [C$_8$mim][PF$_6$].** **(a)** Normalized absorption spectra of TPP in air-saturated ionic liquid [C$_8$mim][PF$_6$]. The inset shows the Q-bands of absorption. **(b)** Transient absorption spectra for TPP instantaneously (at 7 μs) in ionic liquid and neat ionic liquid [C$_8$mim][PF$_6$].

Surprisingly, the dynamics of TPP triplet excited state in the heterogeneously structural ionic liquid [C$_8$mim][PF$_6$] are gradually changed from a fast decay to a slow decay under the air-saturated condition with the increase of laser irradiations from 10 to 300 and up to 500 pulses as shown in Fig. 2a. The dynamics of TPP triplet excited state observed with 10 pulses laser irradiation show a single exponential decay, whereas a biexponential decay law is required to obtain the best fitting after 40 or more pulses. As the laser irradiations increase from 10 to 500 pluses, the average lifetime of TPP triplet state markedly increase from 1.70 to 128.10 μs as listed in Table 1. Furthermore, as a comparative experiment in the deoxygenated ionic liquid [$C_8$mim][$PF_6$], the dynamics of TPP triplet excited state show very slow decays and there is no obvious dynamic changes observed with all the laser irradiations applied from 10 to 500 pulses as shown in Fig. 2b, indicating that oxygen rather than the microstructure of ionic liquid mainly plays the crucial role in modulating the TPP excited triplet state lifetimes. It should be mentioned here from the control experiments that, the exactly same absorption spectra and NMR spectra of TPP in air-saturated ionic liquids before and after laser irradiation as shown in Supplementary Fig. 1 and Supplementary Fig. 3, indicate that there is no structure damage and no reaction between oxygen molecules and TPP or ionic liquids during the laser irradiation. The completely recovery of TPP triplet excited state dynamics after 500 pulses as shown in Supplementary Fig. 4, suggest that the changed TPP triplet state behaviors are possibly from the redistribution of oxygen molecules in the heterogeneous ionic liquids.

To prove that it is the redistributed oxygen in the heterogeneous ionic liquid [$C_8$mim][$PF_6$] that changes the TPP triplet excited state decay after laser irradiation, as comparative experiments, we further perform the transient absorption of TPP triplet excited state in the homogeneously conventional solvents such as polar THF and nonpolar toluene in both air-saturated and deoxygenated conditions as shown in Supplementary Fig. 5. It is found that the triplet state lifetimes of TPP are very short in air-saturated solvents less than 1 μs, and obviously long lifetimes about 150 μs are observed after deoxygenation as listed in Supplementary Table 4. Especially, there are no obviously lifetime changes of TPP triplet state in ether air-saturated or deoxygenated conventional solvents (THF or toluene) with a homogeneous microenvironment no matter how many laser pulses are applied as shown in Supplementary Figs. 5a-d. Since the two conventional solvents have the similar thermal conductivities to ionic liquid [$C_8$mim][$PF_6$], it should be mentioned that the heating effects of the laser pulse irradiations (1 Hz, 7 ns, 10 mJ/pulse, 3mm spot size) for these solvents could be neglected, suggesting that the obvious triplet state lifetime changes mainly result from the oxygen redistribution within the changed heterogeneous microstructure of ionic liquid [$C_8$mim][$PF_6$] after laser irradiation. Furthermore, it is found that the long-lived TPP triplet state dynamics in ionic liquid [$C_8$mim][$PF_6$] as shown in Supplementary Fig. 4c, are completely recovered after about ~16 hours, indicating that these laser-induced behaviors of TPP are mainly related to the highly ordered meso-/microscopic rearrangements of polar and nonpolar domains in non-equilibrium RTILs, rather than the fast structure fluctuation of ionic liquid at equilibrium probed by ultrafast Kerr effect experiments.[61,62] In fact, since the main absorption of ionic liquids are in the region shorter than 300 nm, there is no obvious absorption for ionic liquid itself at laser wavelength (532nm). Photo bleaching/decomposition of ILs may not take place upon laser irradiation, whereas the continuous irradiation by the laser pulses (1 Hz, 7 ns, 10 mJ/pulse) for longer time could be helpful for the formation of highly ordering microstructure of ionic liquids.[25]

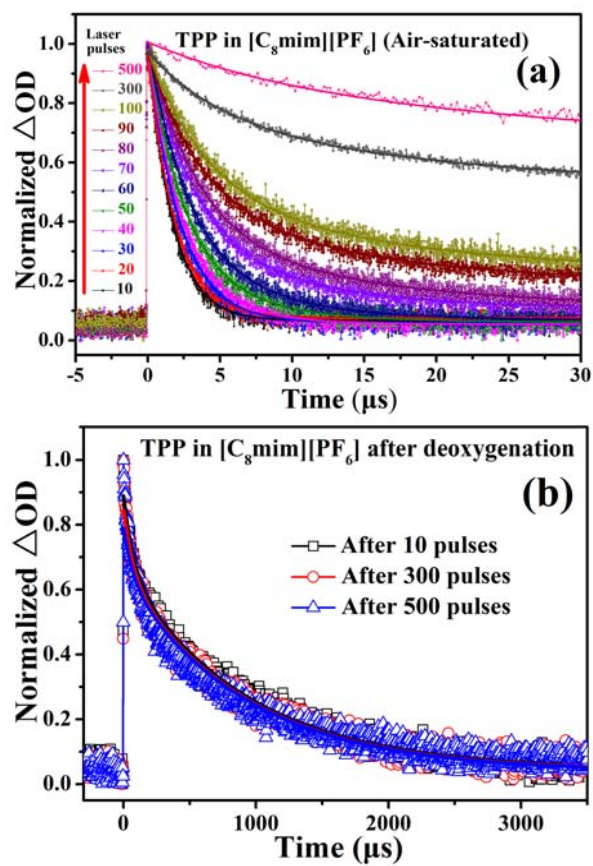

**Figure 2 | Triplet excited state dynamics of TPP in ionic liquid [C$_8$mim][PF$_6$].** Normalized triplet excited state decays of TPP in ionic liquid [C$_8$mim][PF$_6$] by laser flash photolysis from 10 to 500 pulses laser irradiation in the **(a)** air-saturated and **(b)** deoxygenated condition. Excitation at 532 nm; monitored at 450 nm. Fitted curves are shown by solid lines. The value of $\chi^2$ for each fitting was estimated in the range of 0.95-1.05.

**Table 1 | TPP triplet state decay lifetimes after laser irradiations from 10 to 500 pulses in ionic liquid [C$_8$mim][PF$_6$].**

| Laser | Fitted results[a] |
|---|---|
|  |  |

| pulses | $A_1$ | $\tau_1$ (µs) | $A_2$ | $\tau_2$ (µs) | $\tau^b$ (µs) |
|---|---|---|---|---|---|
| 10 | | | | | 1.70 |
| 20 | | | | | 1.87 |
| 30 | | | | | 2.22 |
| 40 | 0.90 | 2.36 | 0.10 | 4.57 | 2.58 |
| 50 | 0.81 | 2.47 | 0.19 | 5.86 | 3.11 |
| 60 | 0.79 | 2.87 | 0.21 | 7.62 | 3.87 |
| 70 | 0.77 | 3.23 | 0.23 | 10.75 | 4.96 |
| 80 | 0.73 | 3.51 | 0.27 | 14.95 | 6.60 |
| 90 | 0.67 | 3.79 | 0.33 | 15.95 | 7.80 |
| 100 | 0.63 | 3.98 | 0.37 | 16.66 | 8.67 |
| 300 | 0.54 | 5.65 | 0.46 | 36.18 | 19.69 |
| 500 | 0.33 | 17.23 | 0.67 | 182.71 | 128.10 |

[a]The single exponential fitting using $r(t) = a_1 e^{-t/\tau_1}$ for 10, 20 and 30 pulses irradiation, respectively, and the biexponential fitting using $r(t) = \sum_{i=1}^{2} a_i e^{-t/\tau_i}$. [b]The average lifetime from biexponential fitting using $<\tau> = \sum_{i=1}^{2} a_i \tau_i / \sum_{i=1}^{2} a_i$, $A_i = a_i / \sum_{i=1}^{2} a_i$, $\sum_{i=1}^{2} A_i = 1$.

In fact, for air-saturated ionic liquid before laser irradiation, the triplet excited state of TPP is mainly quenched by the oxygen molecules presented in polar and nonpolar domains of ionic liquid [C$_8$mim][PF$_6$] with a very fast single exponential decay as shown in Fig. 2a. By the irradiation of laser, the charged imidazolium rings and anions of ionic liquid in the polar domain are likely to arrange and pack themselves along the direction of local laser electric field, and the octyl chains would correspondingly form an ordered nonpolar domain through van der Waals interactions interlaced with the charged polar domain.[29, 30] In this scenarios, both the polar domains compactly formed by the charged components (positive charged imidazolium rings and anions) and nonpolar domains composed of the incompact octyl chains aggregation become more ordered arrangements and integrated along the direction of local laser electric field, leading to the change of heterogeneous microstructure of ionic liquid [$C_8$mim][$PF_6$] upon laser irradiation.

Generally, oxygen molecules are nonpolar molecules, which means the two oxygen atoms have similar electronegativities. However, under external light field (such as laser irradiation) or electric field (such as external electric field, or internal-built ionic field reasonably from integrated polar domain in ionic liquid with positive charged imidazolium rings and anions), electrons can be on one side of the oxygen molecule, so one oxygen atom is slightly electronegative and the other is partially positive, which leads to the electron cloud of a nonpolar oxygen molecule to be displaced in the direction of the external electric field. Therefore, oxygen molecules could be easily polarized by the external laser electric field or by the internal-built ionic field of ionic liquids itself within polar domain and prefer to move close to or enter into the charged polar domains through electrostatic interactions with a more ordered and integrated arrangement under laser irradiation, leading to the redistribution of oxygen in the changed heterogeneous microstructure of ionic liquid with different amount of laser irradiations. As a small molecule, polarized oxygen molecules could more easily diffuse into the compact polar domains composed of charged components after laser irradiation. Meanwhile, the more integrated polar domain containing positive charged imidazolium rings and anions is helpful for withdrawing the polarized oxygen molecules from nonpolar domains to polar domains and keeping polarized oxygen molecules to stay in polar domain due to electrostatic interactions. In this case, the TPP molecules in ionic liquid [$C_8$mim][$PF_6$] would correspondingly experience a heterogeneous microenvironment change with the redistribution of oxygen molecules upon laser irradiation, resulting in lower concentrations of oxygen in nonpolar domains and higher concentrations of oxygen around the charged components of polar domains. As shown in Fig. 2a, the TPP triplet excited state dynamics are sensitively modulated even after several tens pulses laser irradiation in heterogeneous ionic liquid [$C_8$mim][$PF_6$]. The lifetimes of TPP triplet excited state are largely changed in a wide range after different amount of pulses laser irradiation as listed in Table 1. As the oxygen redistributing in the changed heterogeneous microstructure of ionic liquid [$C_8$mim][$PF_6$] after 40 or more pulses of laser irradiation, the decays of TPP triplet excited state dynamics are following a biexponential law with an intensive increase of the slow component up to 500 pulses laser irradiation as shown in Fig. 2 and Table 1, while almost all oxygen molecules diffuse into the polar domain without obvious changes of TPP triplet state dynamics with more than 500 pulses laser irradiation.

In one word, laser irradiation leads to the changes of the heterogeneous microstructure of ionic liquids [$C_8$mim][$PF_6$], which results in the redistribution of oxygen molecules and consequently changing the dynamics of TPP triplet excited state. Since there is no obvious dynamic changes of TPP triplet excited state observed with all the laser irradiations applied from 10 to 500 pulses in the deoxygenated ionic liquid [$C_8$mim][$PF_6$] as well as air-saturated or deoxygenated homogeneous solvents (toluene and THF) as shown in Fig. 2b and Supplementary Figs 5a-d, it is clearly confirmed that it is the redistribution of oxygen molecules in the modulated heterogeneous microstructure of ionic liquid rather than the changed microstructure itself, which causes the changes of the TPP excited triplet state dynamics. Therefore, the remarkably changes of TPP triplet state upon laser irradiation related to the microstructure changes could provide an excellent probe to identify the redistribution of oxygen molecules in heterogeneous ionic liquids. The specific behaviors with microstructure change and redistribution of oxygen in this heterogeneous ionic liquid [$C_8$mim][$PF_6$] upon laser irradiation are described as shown in Fig. 3.

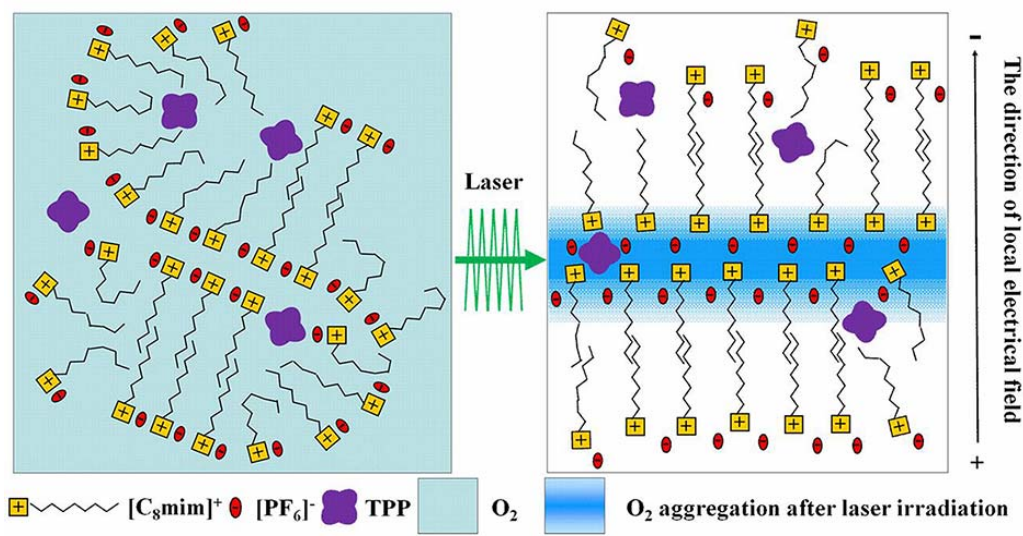

**Figure 3 | Evolution of microstructure and oxygen distribution in heterogeneous ionic liquid [$C_8$mim][$PF_6$] experienced by TPP induced by laser irradiation.** Schematic representation of laser-induced structural heterogeneity changing of ionic liquid [$C_8$mim][$PF_6$] along the local electric field and the redistribution of oxygen experienced by TPP molecules. **|** (Left) before laser irradiation and (right) after laser irradiation.

The redistribution of oxygen after laser irradiations is further confirmed by the molecular dynamics (MD) simulations. The detailed MD methods and MD results are shown in Supplementary Note 5. The polarization of oxygen by laser irradiation results in the charge shift of $O_2$ molecules to form the $O^+$-$O^-$, leading to more charge-shifted $O_2$ redistribution in the polar domains of ionic liquid [C$_8$mim][PF$_6$] than in nonpolar domains. By MD simulations, it is found that polarized oxygen molecules are significantly distributed around the charged components in the polar domain of ionic liquid [C$_8$mim][PF$_6$] as shown in Supplementary Figs 10-12.

**Rotational dynamics of C153.** To prove the heterogeneous microstructure changes, we further perform the study of C153 rotational dynamics in ionic liquid [C$_8$mim][PF$_6$] under different amount of pulses laser irradiation. In our previous study, it is found that the rotational dynamics of C153 are very sensitive to the heterogeneous microstructue of ionic liquids [C$_n$mim][PF$_6$] with different structural domains. The two types of structural domains with different microviscosities are related to the fast and slow rotational dynamics experienced by C153 due to the heterogeneous nature of ionic liquids.[29, 30] Lucky to the very slow recovery (Supplementary Fig. 4c) for structural heterogeneity changes of ionic liquid [C$_8$mim][PF$_6$] after external laser irradiation, we can have chance to measure the TCSPC immediately just right after different amount of pulses laser irradiation for investigating C153 rotational dynamics. The laser-induced changes of heterogeneous microstructure of ionic liquid [C$_8$mim][PF$_6$] probed by C153 rotational dynamics are listed in Table 2 and Supplementary Fig. 13.

For a well dispersed ideal probe of C153, the fast and slow rotation times obtained from C153 rotational dynamics in ionic liquid [C$_8$mim][PF$_6$] are attributed to the nonpolar and polar domains with incompact and compact microstructures, respectively. With the packing of the polar domain after 10 to 500 pulses laser irradiation, more charged components (polar domain) are arranged along the local electric field, making the slight increase of slow component of C153 rotational dynamics from 51% to 58%. This is understood by the competition between the external laser electric field and the internal ionic field through the orientation of the aligned positive charged imidazolium rings and anions related to the laser polarization direction, which will attract more laser-polarized oxygen molecules into polar domains from nonpolar domains. Furthermore, the correspondingly ordered arrangement of octyl chains with the conformation from coiled to extended possibly causes the nonpolar domain more dense with the increase of fast rotation time from 0.47 to 0.65 ns. With all the ordered arrangement of polar and nonpolar domains under laser irradiation, the whole microstructure is more ordered and compacted with the decrease of free volume causing the increase of average rotation time of C153 from 2.84 to 3.25 ns, which is consistent with the previous MD results for ionic liquid under the applied external electric field.[25]

Because the long octyl chains of ionic liquid [$C_8$mim][$PF_6$] will extend and realign along the local laser electric field, the nonpolar domain is possibly larger than the polar domain after laser irradiation, although the microstructure of ionic liquid [$C_8$mim][$PF_6$] becomes more order and compact with a smaller volume upon laser irradiation than the spatially heterogeneous one before irradiation. In this case, the well dispersed TPP within the whole microenvironment of ionic liquid [$C_8$mim][$PF_6$] still mainly stay around the nonpolar domains with long lifetime of their triplet excited state after laser irradiation, while most of polarized oxygen molecules move from nonpolar domains into compact polar domains close to the charged components as the model shown in Fig. 3.

**Table 2 | Rotational relaxation parameters for C153 in ionic liquid [$C_8$mim][$PF_6$] after laser irradiations.**

| | Fitting results[a] | | | | |
|---|---|---|---|---|---|
| Laser pulses | $A_1$ | $\tau_1$ (ns) | $A_2$ | $\tau_2$ (ns) | $\tau_r^{b}$ (ns) |
| 10 | 0.49 | 0.47 | 0.51 | 5.11 | 2.84 |
| 300 | 0.45 | 0.53 | 0.55 | 5.13 | 3.06 |
| 500 | 0.42 | 0.65 | 0.58 | 5.13 | 3.25 |

$^a$The biexponential fitting using $r(t) = \sum_{i=1}^{2} a_i e^{-t/\tau_i}$, $^b$The average lifetime from biexponential fitting using $<\tau_r> = \sum_{i=1}^{2} a_i \tau_i / \sum_{i=1}^{2} a_i$, $A_i = a_i / \sum_{i=1}^{2} a_i$, $\sum_{i=1}^{2} A_i = 1$.

**Alkyl chain-length dependent microstructure changes and redistribution of oxygen molecules in ionic liquids [C$_n$mim][PF$_6$].** In our previous work, we have investigated the heterogeneous microenvironment of a series of ionic liquids [C$_n$mim][PF$_6$] with different alkyl chain lengths ($n$ = 4-8) to examine the essential different microstructures and physical chemistry properties.[30] It is of great importance to have a deep insight into the correlation between the microstructure changes and the redistributions of oxygen behind the effect of molecular structure tailoring with different lengths of side chains on cations upon external laser irradiation. In addition to the above-mentioned results of ionic liquid [C$_8$mim][PF$_6$], for ionic liquids [C$_4$mim][PF$_6$] and [C$_6$mim][PF$_6$] (Scheme 1) taken as examples, it is found that the dynamics of TPP triplet excited state in these air-saturated ionic liquids are significantly different compared to ionic liquid [C$_8$mim][PF$_6$] as shown in Fig. 4 and Supplementary Table 2, but there is no obvious changes observed in all deoxygenated ionic liquids upon laser irradiations, suggesting that the different TPP triplet excited state behavior changes are reasonably resulted by the redistribution of oxygen and closely related to the different heterogeneous microstructures of ionic liquids.

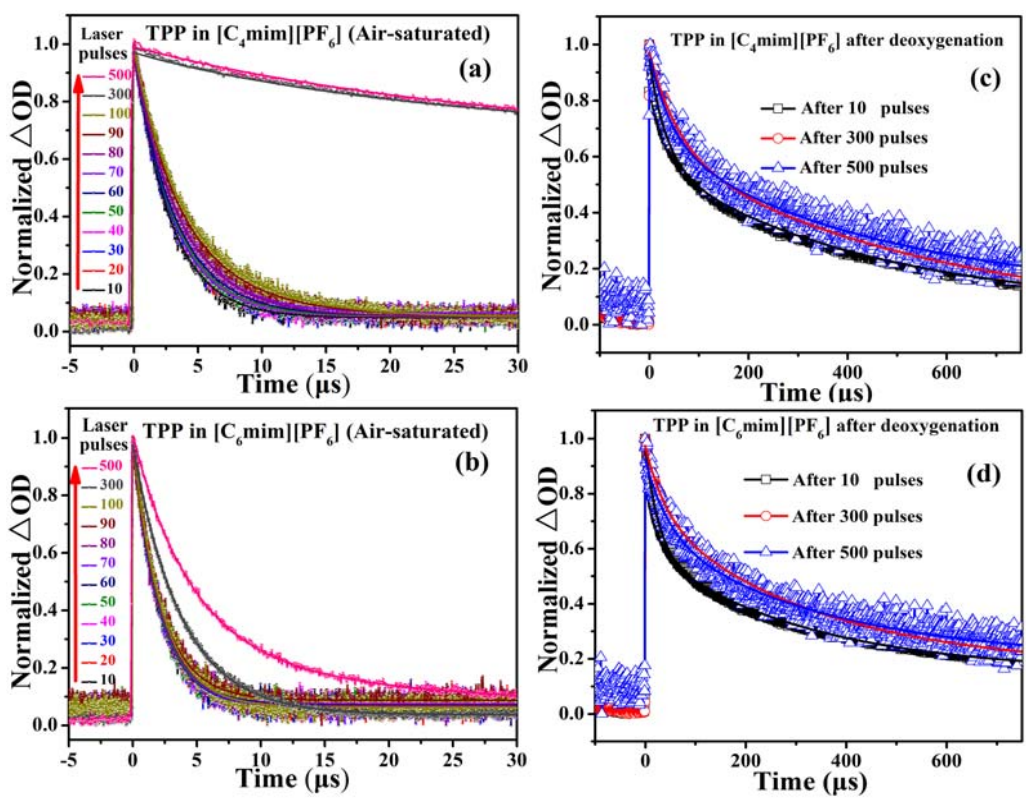

**Figure 4 | TPP Triplet excited state dynamics in ionic liquids [C$_n$mim][PF$_6$] ($n$ = 4, 6).** Normalized TPP triplet excited state decays in **(a) (b)** air-saturated and **(c) (d)** deoxygenated ionic liquids [C$_n$mim][PF$_6$] ($n$ = 4, 6) by laser flash photolysis at 532 nm excitation, monitored at 450 nm. Fitted curves are shown by solid lines. The value of $\chi^2$ for each fitting was estimated in the range of 0.95-1.05.

As shown in Fig. 2a and Fig. 4a, the dynamics of TPP triplet excited state in air-saturated ionic liquid [C$_4$mim][PF$_6$] are less changed with shorter lifetimes than ionic liquid [C$_8$mim][PF$_6$] from 10 to 100 pulses laser irradiation. Due to lack of the local ordered interdigitated components from a shorter butyl chain, the laser-induced microstructure ordering of ionic liquid [C$_4$mim][PF$_6$] along the local laser electric field is less sensitive than ionic liquid [C$_8$mim][PF$_6$] before 100 pulses. As a result, the less integrated polar domains in ionic liquid [C$_4$mim][PF$_6$] hardly keep polarized oxygen molecules for a oxygen redistribution, leading to the less changes of TPP triplet excited state dynamics. The obviously dynamics changes of TPP triplet excited state with long lifetimes in ionic liquid [$C_4$mim][$PF_6$] are observed until about 300 pulses laser irradiation due to the microstructure ordered rearrangement with the help of some extent free volume (Supplementary Fig. 13a) along with the redistribution of oxygen.

In addition, the components proportions of TPP triplet excited state dynamics are almost unchanged in ionic liquid [$C_4$mim][$PF_6$] after 300 pulses laser irradiation, suggesting that its structural heterogeneity rearrangement has almost completed and reached a relatively steady status after 300 laser pulses. This is in agreement with the result of C153 rotational behaviors for [$C_4$mim][$PF_6$] as shown in Supplementary Fig. 13 and Supplementary Table 5, where the proportion of the slow component of C153 rotational dynamics increases from initial 51% at 10 pulses to 58% at 300 pulses and 57% at 500 pulses, respectively, indicating that the orderly integrated polar domains are almost completely formed after 300 pulses laser irradiation. As shown in Supplementary Fig. 4 and Supplementary Table 3, the recovery time of ionic liquid [$C_4$mim][$PF_6$] with a shorter butyl chain after 500 pulses irradiation is relatively short about 6 hours; whereas the recovery time of ionic liquid [$C_8$mim][$PF_6$] is very long up to about 16 hours after 500 pulses irradiation due to more ordered microstructrue with interdigitated longer octyl chains.

Unexpectedly, the lifetime of TPP triplet excited state in air-saturated ionic liquid [$C_6$mim][$PF_6$] has almost no obviously change as the laser irradiation up to 300 pulses, and only a slightly longer lifetime is observed up to 500 pulses laser irradiation with a fast recovery time only about 3 hours (Supplementary Fig. 4 and Supplementary Table 3), because there is no enough free volume for the laser-induced heterogeneous microstructure rearrangement and the redistribution of oxygen in ionic liquid [$C_6$mim][$PF_6$]. This can be also proved by the results of C153 rotational dynamics in ionic liquid [$C_6$mim][$PF_6$], where the rotational behaviors of C153 are almost unchanged even after laser irradiations with 300 and 500 pulses, suggesting no obvious microstructure changes as shown in Supplementary Fig. 13 and Supplementary Table 5. This indicates that there is almost no free volume from the heterogeneous microstructure of ionic liquid [$C_6$mim][$PF_6$] with hexyl chains different from ionic liquid with shorter butyl chain [$C_4$mim][$PF_6$] or longer octyl chain [$C_8$mim][$PF_6$] as shown in Supplementary Fig. 13a. In this case, the heterogeneous microstructure with nonpolar and polar domains for ionic liquid [$C_6$mim][$PF_6$] has no obvious changes, and the redistribution of oxygen molecules will not obviously occur in ionic liquid [$C_6$mim][$PF_6$] even upon the external laser irradiation.

**Discussion**

We have presented a microstructure-dependent modulation of porphyrin triplet excited state dynamics in the controllable RTILs heterogeneous microenvironment with the redistribution of oxygen by the external laser irradiations. The laser-induced microstructure changes with the redistribution of oxygen in ionic liquids are also supported by the results of C153 rotational dynamics. The polarization of oxygen by laser irradiation results in nonpolar $O_2$ molecules to form the $O^+$-$O^-$, leading to more charge-shifted $O_2$ move into polar domains from nonpolar domains in ionic liquid [$C_8$mim][$PF_6$]. The polar domains can keep polarized oxygen molecules to stay in polar domain due to electrostatic interactions, while the polarized $O_2$ molecules gradually move out from polar region to nonpolar region in ionic liquid with the recovery of microstructure after laser irradiation is off. It is found that the dynamics of porphyrin triplet states can be sensitively affected in air-saturated ionic liquids, and a remarkably long lifetime of the triplet excited state are obtained after external laser irradiation, while the redistribution of oxygen molecules from the microstructure changes of ionic liquid plays the crucial role in changing the TPP triplet excited state dynamics.

Furthermore, the redistribution of $O_2$ in ionic liquids upon external field stimuli should be a common phenomenon for many other similar molecules such as $CO_2$ and $CS_2$, etc.. As designer solvents of RTILs, their heterogeneous microstructure can be manipulated for numerous possibilities upon external field stimuli, and the related responsive modes of these heterogeneous systems by external laser irradiation could be significantly different even with only a slightly change of RTILs molecular structure. Therefore, combined with various solutes through external stimuli like laser irradiation or electric field, these heterogeneous systems can be designed and possibly bring extensive inspirations for those chemical reactions associated with not only for $O_2$ but also for $CO_2$, $CS_2$ and many other similar molecules or solvents in many heterogeneous reaction systems, such as catalysis, lithium-air batteries, gaseous sensing and bio-reaction,etc..

## Methods

**Materials.** $[C_n\text{mim}][PF_6]$ ($n$ = 4, 6, 8) ionic liquids (purity >99%, water content <500 ppm) were purchased from Lanzhou Greenchem ILS, LICP, CAS, China, and stored under a dry nitrogen atmosphere. Deuterated dimethyl sulfoxide (DMSO-$d_6$) (purity >98%, with 0.03% TMS) for NMR was from Innochem, China. Tetrahydrofuran (THF) and toluene used in this work were of analytical reagent grade and were purchased from the Beijing Chemical Plant. TPP and C153 were from Fluka and Aldrich, respectively, and used as received. In order to avoid the aggregation, the concentrations of TPP and C153 in all related samples are low as about $3.5 \times 10^{-6}$ and $1.25 \times 10^{-5}$ M, respectively.

**Steady-state absorption/emission spectroscopy.** Ultraviolet/visible (UV/vis) absorption and fluorescence spectra were measured on a spectrophotometer (U3010, Hitachi, Japan) and a fluorescence spectrometer (F4600, Hitachi, Japan), respectively.

**Laser flash photolysis after different amount of pulses laser irradiations.** Nanosecond time-resolved transient absorption (ns-TA) spectra and triplet excited state decay dynamics were measured by a nanosecond laser flash photolysis (LFP) setup Edinburgh LP920 spectrometer (Edinburgh Instruments Ltd.), combined with a Nd:YAG laser (Surelite Ⅱ, Continuum Inc.). The same laser at 532nm is used for both laser irradiation treatment with different amount of laser pulses and excitation for laser flash photolysis experiments. The laser spot size is about ~3 mm in diameter. Since 10 laser pulses are required for each ns-TA measurement, the TPP triplet excited state dynamics at the initial conditions in ionic liquids are obtained with 10 pulses laser irradiations. The excitation wavelength is 532 nm laser pulse (1 Hz, 10 mJ/pulse, fwhm ≈ 7 ns). The analyzing light was from 450 W pulsed xenon lamp. A monochromator equipped with a photomultiplier for collecting the spectral range from 350 to 800 nm was used to analyze transient absorption spectra. All the samples were freshly prepared and well sealed for each measurement in a 10 mm path length quartz cuvettes. Data were analyzed by the online software of the LP920 spectrophotometer. The fitting quality was judged by weighted residuals and reduced $\chi^2$ value.

**Nuclear magnetic resonance spectroscopy (NMR).** The stabilities of air-saturated ionic liquids [$C_n$mim][$PF_6$] ($n$ = 4, 6, 8) containing TPP (i.e. ionic liquids + TPP) before and after laser irradiations are characterized by $^1$H NMR spectra using Deuterated dimethylsulfoxide (DMSO-d6) as a solvent. $^1$H NMR spectra were obtained on a Bruker Avance-400 Ⅲ (400 MHz) spectrometer.

**Time-resolved fluorescence anisotropy measurements after different amount of pulses laser irradiations.** The time-resolved fluorescence anisotropy measurements were carried out using a time-correlated single-photon counting (TCSPC) spectrometer (F900, Edinburgh Instrument) as described before.[29, 30] The samples were excited at 450 nm using a picosecond LED source (PLS-500, PicoQuant, Germany). The instrument response function (IRF) of the detection system is less than 300 ps. The time-resolved fluorescence anisotropy experiments were carried out immediately just right after different amount of pulses laser irradiation treatments to the samples by using a 7 ns YAG laser (Surelite Ⅱ, Continuum Inc.) with 532 nm, 1Hz, 10 mJ/pulse (the size of laser irradiation spot on sample is about ~3 mm during irradiation). The fluorescence anisotropy dynamics were measured right from the YAG laser irradiation region in samples, and the polarization directions of both excitation laser for TCSPC and the YAG laser for irradiation are arranged in parallel during irradiation and lifetime measurements.

## Acknowledgment

This work was supported by the 973 Program (No. 2013CB834604), NSFCs (Nos. 21333012, 21673252 and 21373232) and the Strategic Priority Research Program of the Chinese Academy of Sciences (Grant No. XDB12020200).

## Author contributions

R.M., B.L. and A.X. conceived the research, R.M., B.L., X.W., J.J., Z.K. and Q.G. performed the experiments, L.Y. performed the MD calculation, R.M. and B.L. analysed the results, R.M., B.L. and A.X. drafted the manuscript. All authors revised and proof read the paper.

## Additional Information

**Supplementary Information** accompanies this paper at

http://www.nature.com/naturecommunications. Detail steady-state and transient-state spectroscopy measurements, Nuclear magnetic resonance spectroscopy (NMR) measurement, molecular dynamics (MD) simulation, rotational dynamics of coumarin 153.

**Competing financial interests: The authors declare no competing financial interests.**

**TOC Graphic**

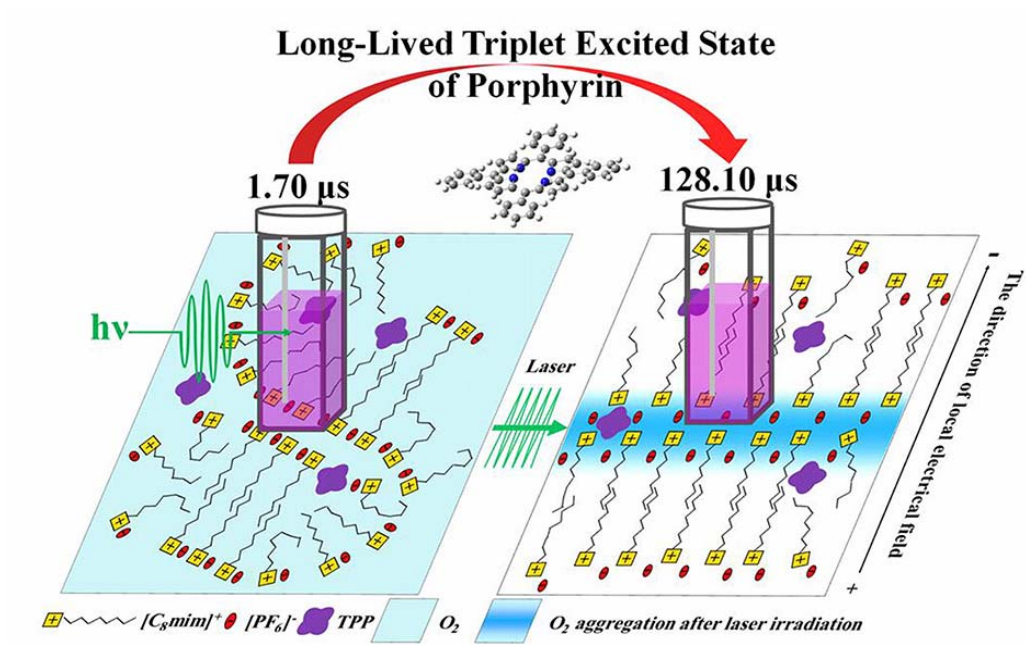

# Supporting Information

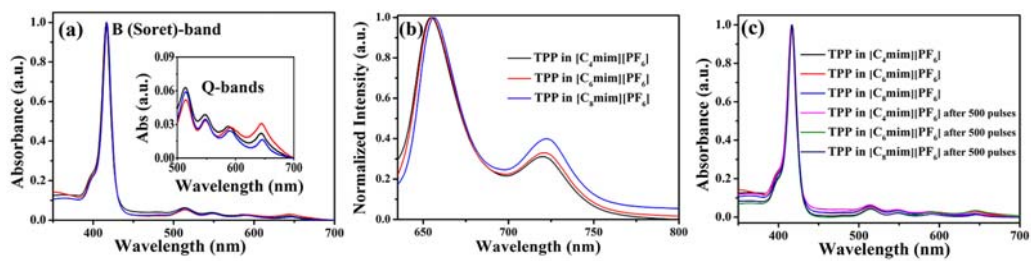

**Supplementary Figure 1.** Normalized (a) absorption and (b) fluorescence spectra of TPP in ionic liquids [C$_n$mim][PF$_6$] ($n$ = 4, 6, 8) (before laser irradiation). The inset shows the Q-bands of absorption. Excitation wavelength for fluorescence spectra is at 532 nm. (c) Normalized absorption spectra of TPP in ionic liquids [C$_n$mim][PF$_6$] ($n$ = 4, 6, 8) before and after 500 pulses laser irradiation. This indicates that no change or photodamage for TPP after laser irradiations.

**Supplementary Table 1. Steady-state absorption and fluorescence properties of TPP in ionic liquids [C$_n$mim][PF$_6$] ($n$ = 4, 6, 8) before laser irradiation.**

|  | $\lambda_{Abs}^{a}$ (nm) | $\lambda_{Em}^{b}$ (nm) | $\Delta\lambda^{c}$ (nm) |
|---|---|---|---|
| [C$_4$mim][PF$_6$] | 416 | 654 | 238 |
| [C$_6$mim][PF$_6$] | 417 | 655 | 238 |
| [C$_8$mim][PF$_6$] | 417 | 656 | 239 |

$^a$The maxima absorption peak wavelength. $^b$The maxima fluorescence emission peak wavelength. $^c$The Stokes shift.

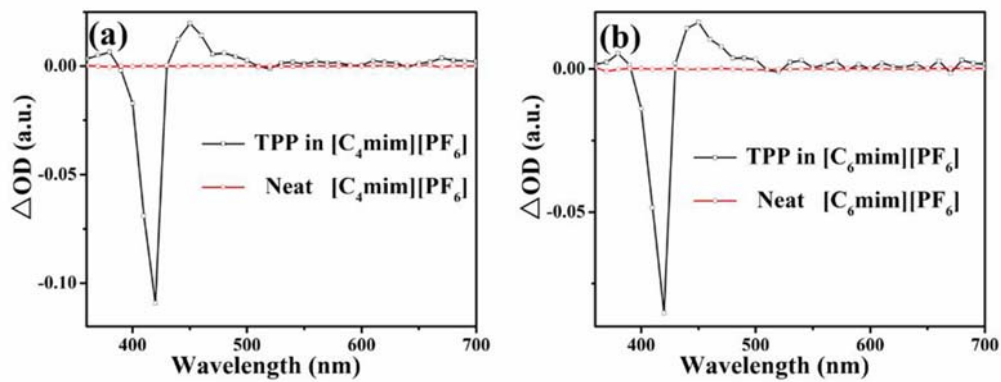

**Supplementary Figure 2.** Transient absorption spectra of TPP in ionic liquids and neat ionic liquids [$C_n$mim][$PF_6$] ($n$ = 4, 6) before laser irradiation, at delay times (7 μs), 532 nm excitation by laser flash photolysis.

## Supplementary Note 1

**Triplet Excited State Behavior of TPP in Air-saturated and Deoxygenated Ionic Liquids [$C_n$mim][$PF_6$] after Laser Irradiation.**

In air-saturated ionic liquids [$C_n$mim][$PF_6$] ($n$ = 4, 6), the TPP triplet excited state dynamics upon different laser pulse irradiations are carefully studied. As listed in Supplementary Table 2, with 10 pulses/step increase, the lifetime of TPP triplet state in ionic liquid [$C_4$mim][$PF_6$] becomes longer from 2.79 to 3.67 μs after laser irradiation from 10 to 80 pulses by a single exponential fitting, and further become longer to an average lifetime from 4.16 to 4.62 μs fitted with a biexponential decay with the laser pulse increases from 90 to 100. After 300 pulses laser irradiation, the triplet state decay of TPP becomes obviously slow followed by a biexponential law, with a longer triplet excited state lifetime of 15.88 μs (29%) and a markedly longer lifetime of 119.60 μs (71%). The proportions of the two components of TPP triplet state decay dynamics are almost unchanged after laser irradiation from 300 to 500 pulses, indicating that the rearrangement of ionic liquid [$C_4$mim][$PF_6$] has almost possibly completed after 300 pulses, which can be further proved by C153 in later section. As the oxygen molecules completely move into the polar domain of [$C_4$mim][$PF_6$] upon the laser irradiation pulses from 300 to 500, the TPP triplet state decay becomes slower with a longer average lifetime from 89.52 to 127.74 μs.

It is found that almost all the triplet excited state dynamics of TPP follow a single exponential decay in ionic liquid [$C_6$mim][$PF_6$], and the lifetime is unchanged from 10 to 50 pulses of laser irradiation and slightly becomes longer from 2.07 to 2.27 μs with the laser pulses increases from 50 to 100. The lifetime of TPP triplet excited state in ionic liquid [$C_6$mim][$PF_6$] changes only in a small extent from 2.07 (10 laser pulses) to 3.75 μs (300 laser pulses) and to an average lifetime of 7.85 μs (500 laser pulses). The triplet state decay of TPP follows a biexponential law until after 500 pulses.

The triplet excited state dynamics of TPP are more sensitively changed in ionic liquid [$C_8$mim][$PF_6$] from a single exponential decay to a biexponential decay after only 10 to 30 pulses of laser irradiation with a longer lifetime from 1.70 to 2.22 μs. With the laser irradiation pulse further increases from 40 to 100, the fast lifetime ($\tau_1$) of TPP triplet state becomes longer from 2.36 to 3.98 μs with its component proportion decreasing from 90% to 63%, while the slow lifetime ($\tau_2$) obviously becomes longer from 4.57 to 16.66 μs with its component proportion increasing from 10% to 37%. The average lifetime of TPP triplet state becomes longer from 2.58 to 8.67 μs. With the laser irradiations up to 300 and 500 pulses, the dynamics of TPP follow a biexponential law with a longer average lifetime of 19.69 μs and a significantly longer one of 128.10 μs, respectively. The two corresponding lifetimes of TPP triplet excited state decays become longer (from 5.65 to 17.23 μs for $\tau_1$; from 36.18 to 182.71 μs for $\tau_2$) and the proportion of slow lifetime ($\tau_2$) component increases from 46% to 67%. The heterogeneous microstructure of ionic liquid [$C_8$mim][$PF_6$] is obviously changed even after 300 to 500 pulses laser irradiation, which is also in agreement with the results of C153 rotation dynamics. With a long alkyl chain for a possibly larger volume of nonpolar domain than polar domain in ionic liquid [$C_8$mim][$PF_6$], the observed triplet excited state dynamics of well dispersed TPP molecules are not mainly affected by the polar domain.

After deoxygenation for ionic liquids [$C_n$mim][$PF_6$] ($n$ = 4, 6), the dynamics of TPP excited triplet state are just slightly changed by laser irradiations, and the results are similar to the experiments when ionic liquid [$C_8$mim][$PF_6$] was used as the solvent as shown in Fig. 2 in main text, which shows that the redistribution of oxygen molecules within the heterogeneous microstructure of ionic liquids are the key factor for modulating the TPP excited triplet state dynamics.

**Supplementary Table 2. TPP triplet state decay lifetimes after laser irradiations of 10 to 500 pulses in air-saturated and deoxygenated ionic liquids [$C_n$mim][$PF_6$] ($n$ = 4, 6, 8)**

| Ionic Liquids [$C_n$mim][$PF_6$] | Laser pulses for irradiation | Fitting results[a] (air-saturated) | | | | |
|---|---|---|---|---|---|---|
| | | $A_1$ | $\tau_1$ (μs) | $A_2$ | $\tau_2$ (μs) | $\tau^{b}$ (μs) |
| [$C_4$mim][$PF_6$] | After 10 pulses | | | | | 2.79 |
| | After 20 pulses | | | | | 3.03 |
| | After 30 pulses | | | | | 3.04 |
| | After 40 pulses | | | | | 3.05 |
| | After 50 pulses | | | | | 3.16 |
| | After 60 pulses | | | | | 3.32 |
| | After 70 pulses | | | | | 3.54 |
| | After 80 pulses | | | | | 3.67 |
| | After 90 pulses | 0.38 | 2.61 | 0.62 | 5.11 | 4.16 |
| | After 100 pulses | 0.47 | 2.93 | 0.53 | 6.11 | 4.62 |
| | After 300 pulses | 0.29 | 15.88 | 0.71 | 119.60 | 89.52 |
| | After 500 pulses | 0.32 | 21.56 | 0.68 | 177.70 | 127.74 |
| [$C_6$mim][$PF_6$] | After 10 pulses | | | | | 2.07 |
| | After 20 pulses | | | | | 2.07 |
| | After 30 pulses | | | | | 2.07 |
| | After 40 pulses | | | | | 2.07 |
| | After 50 pulses | | | | | 2.07 |
| | After 60 pulses | | | | | 2.16 |
| | After 70 pulses | | | | | 2.16 |
| | After 80 pulses | | | | | 2.17 |
| | After 90 pulses | | | | | 2.23 |

|  | | $A_1$ | $\tau_1$ (μs) | $A_2$ | $\tau_2$ (μs) | $\tau^b$ (μs) |
|---|---|---|---|---|---|---|
| | After 100 pulses | | | | | 2.27 |
| | After 300 pulses | | | | | 3.75 |
| | After 500 pulses | 0.80 | 4.53 | 0.20 | 21.11 | 7.85 |
| | Fitting results[a] (deoxygenated) | | | | | |
| | | $A_1$ | $\tau_1$ (μs) | $A_2$ | $\tau_2$ (μs) | $\tau^b$ (μs) |
| [C$_4$mim][PF$_6$] | After 10 pulses | 0.36 | 53.13 | 0.64 | 431.40 | 297.06 |
| | After 300 pulses | 0.34 | 50.83 | 0.66 | 460.43 | 321.17 |
| | After 500 pulses | 0.33 | 46.20 | 0.67 | 459.33 | 323.00 |
| [C$_6$mim][PF$_6$] | After 10 pulses | 0.50 | 28.98 | 0.50 | 380.30 | 204.64 |
| | After 300 pulses | 0.42 | 51.06 | 0.58 | 381.13 | 242.51 |
| | After 500 pulses | 0.38 | 47.92 | 0.62 | 355.71 | 238.75 |
| [C$_8$mim][PF$_6$] | After 10 pulses | 0.35 | 75.01 | 0.65 | 915.23 | 621.15 |
| | After 300 pulses | 0.36 | 75.32 | 0.64 | 921.76 | 617.49 |
| | After 500 pulses | 0.35 | 75.84 | 0.65 | 925.61 | 628.41 |

[a]The single exponential fitting using $r(t) = a_1 e^{-t/\tau_1}$, The biexponential fitting using $r(t) = \sum_{i=1}^{2} a_i e^{-t/\tau_i}$, [b]The average lifetime from biexponential fitting using $<\tau> = \sum_{i=1}^{2} a_i \tau_i / \sum_{i=1}^{2} a_i$, $A_i = a_i / \sum_{i=1}^{2} a_i$, $\sum_{i=1}^{2} A_i = 1$.

## Supplementary Note 2

**NMR Characterization of TPP in Air-Saturated Ionic liquids [C$_n$mim][PF$_6$] ($n$ = 4, 6, 8) Before and After Laser Irradiation.**

The stabilities of air-saturated ionic liquids [C$_n$mim][PF$_6$] ($n$ = 4, 6, 8) containing TPP (i.e. ionic liquids + TPP) before and after laser irradiations are characterized by $^1$H NMR spectra using Deuterated dimethylsulfoxide (DMSO-d6) as a solvent as shown in Supplementary Fig. 3. The $^1$H NMR samples after 500 laser pulses are obtained right from the position of laser irradiation spot (laser: 7 ns YAG laser (Surelite Ⅱ, Continuum Inc.) with 532 nm, 1Hz, 10 mJ/pulse). The samples without laser irradiation are also characterized by $^1$H NMR as a control. Because of the low concentration of TPP about $3.5 \times 10^{-6}$ M, the NMR spectra are mainly from the ionic liquids [C$_n$mim][PF$_6$]. TPP in [C$_4$mim][PF$_6$], $^1$H NMR (400 MHz, DMSO) δ 9.09 (s, 1H), 7.76 (s, 1H), 7.69 (s, 1H), 4.15 (t, $J$ = 7.2 Hz, 2H), 3.84 (s, 3H), 1.76 (m, 2H), 1.26 (m, 2H), 0.91 (t, $J$ = 7.4 Hz, 3H). TPP in [C$_4$mim][PF$_6$] after 500 pluses, $^1$H NMR (400 MHz, DMSO) δ 9.09 (s, 1H), 7.76 (s, 1H), 7.69 (s, 1H), 4.15 (t, $J$ = 7.2 Hz, 2H), 3.84 (s, 3H), 1.76 (m, 2H), 1.26 (m, 2H), 0.90 (t, $J$ = 7.4 Hz, 3H). TPP in [C$_6$mim][PF$_6$], $^1$H NMR (400 MHz, DMSO) δ 9.09 (s, 1H), 7.76 (s, 1H), 7.69 (s, 1H), 4.14 (t, $J$ = 7.2 Hz, 2H), 3.84 (s, 3H), 1.77 (m, 2H), 1.25 (m, 6H), 0.86 (t, $J$ = 6.8 Hz, 3H). TPP in [C$_6$mim][PF$_6$] after 500 pluses, $^1$H NMR (400 MHz, DMSO) δ 9.09 (s, 1H), 7.76 (s, 1H), 7.69 (s, 1H), 4.15 (t, $J$ = 7.2 Hz, 2H), 3.84 (s, 3H), 1.77 (m, 2H), 1.25 (m, 6H), 0.86 (t, $J$ = 6.7 Hz, 3H). TPP in [C$_8$mim][PF$_6$], $^1$H NMR (400 MHz, DMSO) δ 9.09 (s, 1H), 7.76 (s, 1H), 7.69 (s, 1H), 4.14 (t, $J$ = 7.2 Hz, 2H), 3.84 (s, 3H), 1.77 (m, 2H), 1.25 (s, 10H), 0.86 (t, $J$ = 6.7 Hz, 3H). TPP in [C$_8$mim][PF$_6$] after 500 pluses, $^1$H NMR (400 MHz, DMSO) δ 9.09 (s, 1H), 7.76 (s, 1H), 7.69 (s, 1H), 4.14 (t, $J$ = 7.2 Hz, 2H), 3.84 (s, 3H), 1.77 (m, 2H), 1.25 (s, 10H), 0.86 (t, $J$ = 6.7 Hz, 3H).

Both the chemical shift and proton integral of NMR spectra for the ionic liquids [C$_n$mim][PF$_6$] systems with TPP (i.e. ionic liquids + TPP) remain the same before and after 500 pulses laser irradiations, which indicates the systems of TPP in air-saturated ionic liquids [C$_n$mim][PF$_6$] ($n$ = 4, 6, 8) are stable during laser irradiation.

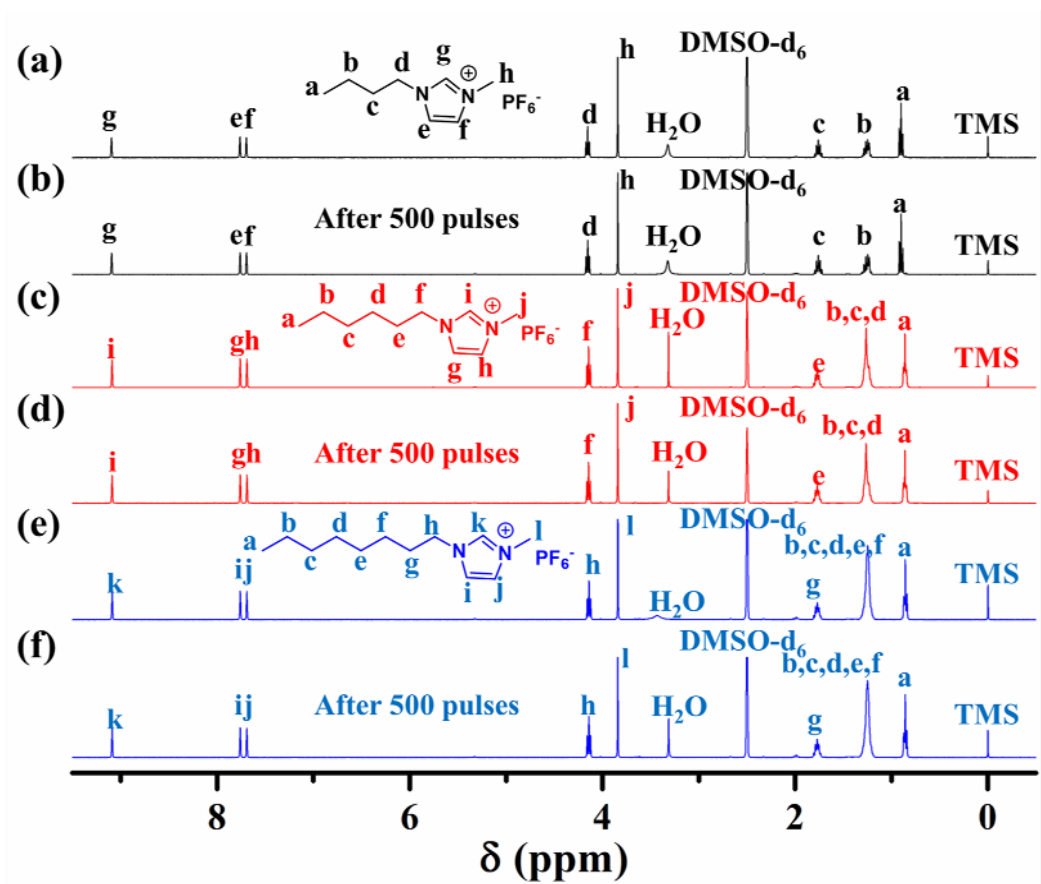

**Supplementary Figure 3.** $^1$H NMR spectra of the systems for TPP in air-saturated ionic liquids [C$_n$mim][PF$_6$] (*n* = 4, 6, 8) before and after 500 laser pulses. (a) TPP in [C$_4$mim][PF$_6$] before laser irradiation, (b) TPP in [C$_4$mim][PF$_6$] after 500 laser pulses; (c) TPP in [C$_6$mim][PF$_6$] before laser irradiation, (d) TPP in [C$_6$mim][PF$_6$] after 500 laser pulses; (e) TPP in [C$_8$mim][PF$_6$] before laser irradiation, (f) TPP in [C$_8$mim][PF$_6$] after 500 laser pulses.

## Supplementary Note 3

**Triplet Excited State Behavior of TPP with the Recovery of the Heterogeneous Microstructure of Ionic Liquids [C$_n$mim][PF$_6$] after 500 Pulses Laser Irradiation.**

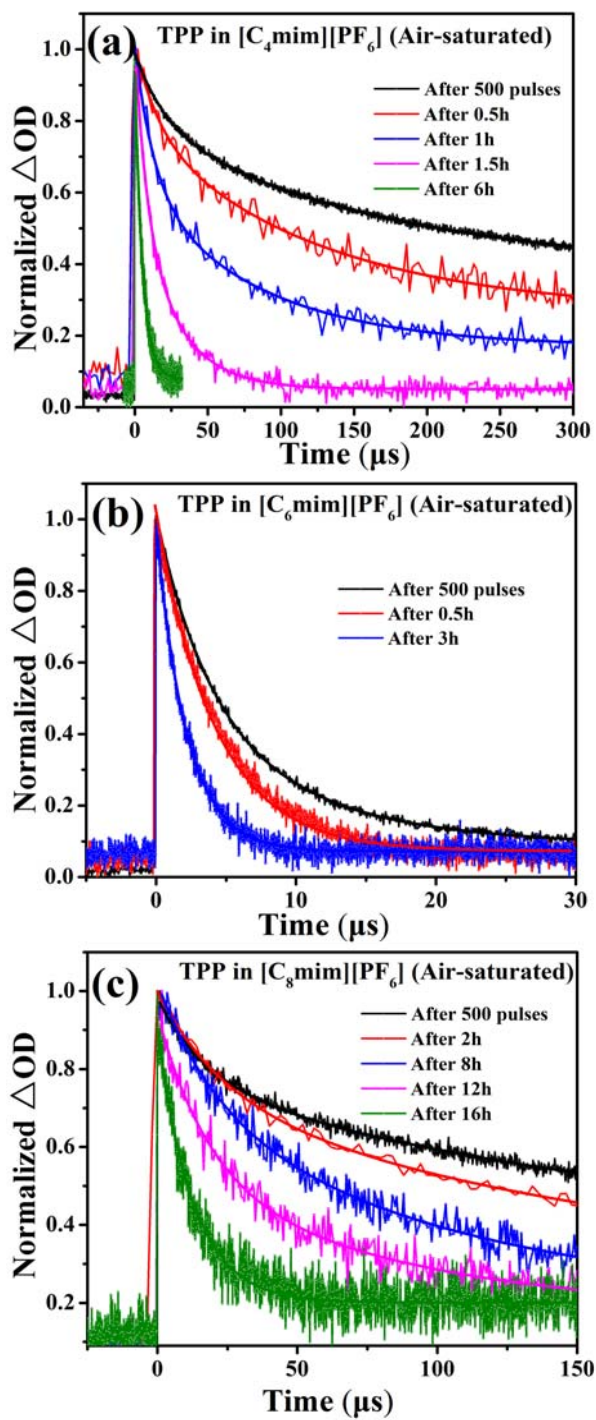

**Supplementary Figure 4.** Normalized triplet excited state decays of TPP in ionic liquids (a) [C$_4$mim][PF$_6$], (b) [C$_6$mim][PF$_6$], and (c) [C$_8$mim][PF$_6$] by laser flash photolysis with different recovery time after 500 pulses laser irradiation. Excitation at 532 nm; monitored at 450 nm. Fitted curves are shown by solid lines. It is found that all the dynamics are recovered.

**Supplementary Table 3. TPP triplet state decay lifetimes with different recovery time after laser irradiation with 500 pulses in ionic liquids [$C_n$mim][$PF_6$] ($n$ = 4, 6, 8)**

| | Time after 500 pulses irradiation | Fitting results[a] | | | | |
|---|---|---|---|---|---|---|
| | | $A_1$ | $\tau_1$ (μs) | $A_2$ | $\tau_2$ (μs) | $\tau$[b] (μs) |
| [$C_4$mim][$PF_6$] | 0.0 hour after 500 pulses | 0.32 | 21.56 | 0.68 | 177.70 | 127.74 |
| | 0.5 hours | 0.33 | 15.97 | 0.67 | 99.32 | 71.81 |
| | 1.0 hours | 0.47 | 12.95 | 0.53 | 82.71 | 49.92 |
| | 1.5 hours | 0.58 | 7.75 | 0.42 | 26.97 | 15.82 |
| | 6.0 hours | | | | | 5.33 |
| [$C_6$mim][$PF_6$] | 0.0 hour after 500 pulses | 0.80 | 4.53 | 0.20 | 21.11 | 7.85 |
| | 0.5 hours | | | | | 4.20 |
| | 3.0 hours | | | | | 2.17 |
| [$C_8$mim][$PF_6$] | 0.0 hour after 500 pulses | 0.33 | 17.23 | 0.67 | 182.71 | 128.10 |
| | 2.0 hours | 0.34 | 16.02 | 0.66 | 160.69 | 111.50 |
| | 8.0 hours | 0.38 | 15.64 | 0.62 | 93.64 | 64.00 |
| | 12.0 hours | 0.55 | 14.02 | 0.45 | 79.25 | 43.37 |
| | 16.0 hours | | | | | 12.77 |

[a]The single exponential fitting using $r(t) = a_1 e^{-t/\tau_1}$, The biexponential fitting using $r(t) = \sum_{i=1}^{2} a_i e^{-t/\tau_i}$, [b]The average lifetime from biexponential fitting using $<\tau> = \sum_{i=1}^{2} a_i \tau_i / \sum_{i=1}^{2} a_i$, $A_i = a_i / \sum_{i=1}^{2} a_i$, $\sum_{i=1}^{2} A_i = 1$.

As shown in Fig. S4, the dynamics of TPP triplet excited state in air-saturated ionic liquids [$C_n$mim][$PF_6$] are changed after 500 laser pulses irradiation with different recovery time. The laser-induced heterogeneous microstructures of ionic liquids along the local electric field gradually recovers to their initial conditions related to the distribution of oxygen at different recovery times. According to the lifetime changes fitted from TPP triplet excited state decays, it is found that, the recovery time of ionic liquid [C$_4$mim][PF$_6$] is about 6 hours with a short butyl side chain, whereas the recovery time of ionic liquid [C$_6$mim][PF$_6$] is much shorter only about 3 hours. For ionic liquid [C$_8$mim][PF$_6$], the orderly heterogeneous microstructure induced by laser irradiation could possibly be more stable with the help of the originally interdigitated octyl chains, and the recovery time of ionic liquid [C$_8$mim][PF$_6$] is relatively longer about 16 hours.

## Supplementary Note 4

**Triplet Excited State Behavior of TPP in Conventional Solvents**

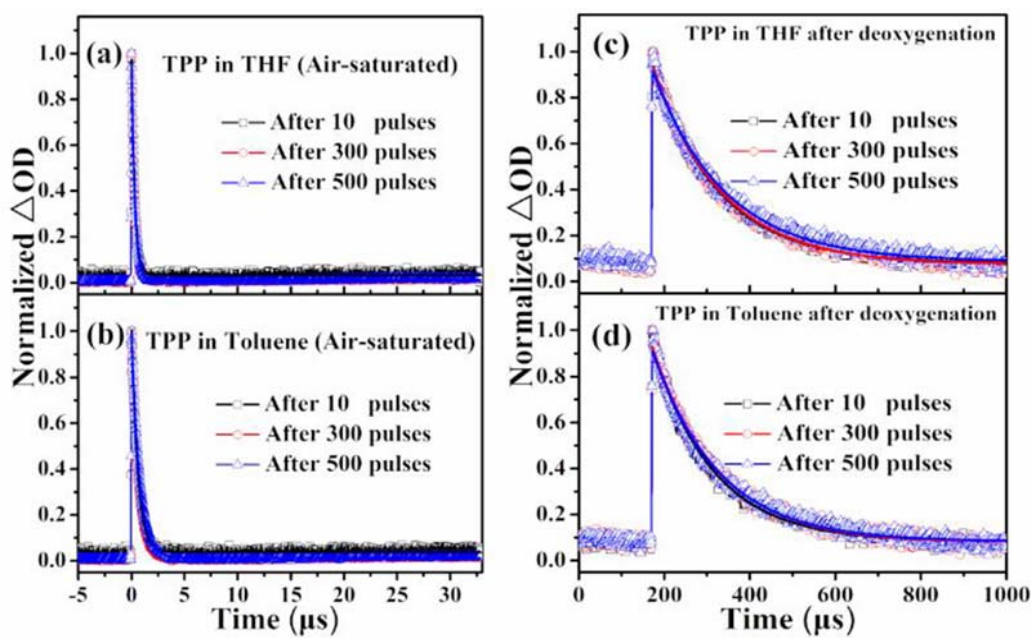

**Supplementary Figure 5.** Normalized triplet excited state decays of TPP in (a) (b) conventional solvents (air-saturated) and (c) (d) in the conventional solvents after deoxygenation by laser flash photolysis from 10 to 500 pulses laser irradiation. Excitation at 532 nm; monitored at 450 nm. Fitted curves are shown by solid lines.

**Supplementary Table 4. TPP triplet decay lifetimes in air-saturated and deoxygenated conventional solvents after laser irradiations**

| Solvents | Laser pulses for irradiation | Fitting results[a] | |
| --- | --- | --- | --- |
| | | $\tau$ (μs) (air-saturated) | $\tau$ (μs) (deoxygenated) |
| Toluene | After 10 pulses | 0.31 | 153.99 |
| | After 300 pulses | 0.30 | 161.27 |
| | After 500 pulses | 0.30 | 168.13 |
| THF | After 10 pulses | 0.57 | 141.95 |
| | After 300 pulses | 0.60 | 153.19 |
| | After 500 pulses | 0.73 | 153.09 |

[a]The single exponential fitting using $r(t) = a_1 e^{-t/\tau_1}$.

It should be noted here that the thermal conductivities of the three ionic liquids [C$_n$mim][PF$_6$] ($n$ = 4, 6, 8) and two compared conventional solvents toluene (nonpolar) and tetrahydrofuran (THF) (polar) are almost the same.[1,2] While the triplet excited state dynamics of TPP in the two of conventional solvents are almost unchanged after 10, 300, and 500 pulses in air-saturated or deoxygenated condition, suggesting that the heating effects by laser irradiation with low repetition (1 Hz) in our experiment for the investigation of the dynamics changes of TPP triplet excited state in ionic liquids [C$_n$mim][PF$_6$] induced by laser could be neglected even if there are any.

## Supplementary Note 5

**MD Simulation of Ionic (Polar) and Hydrophobic (Nonpolar) Domains in the Ionic Liquid.**

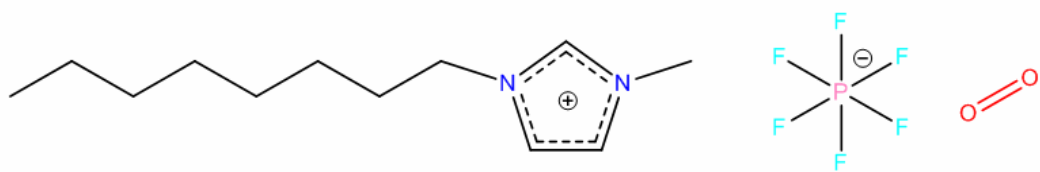

**Supplementary Scheme 1.** Molecular structures of [C$_8$mim]$^+$, [PF$_6$]$^-$ and the molecule of oxygen.

In the following figs, different colors represent different elements. H atoms are white, C atoms are gray, N atoms are blue, O atoms are red, F atoms are cyan and P atoms are pink.

Molecular dynamic (MD) simulation has been successfully used to simulate the ionic and hydrophobic domains in ionic liquids.[3-5] In order to identify the distribution of oxygen molecules in the ionic (polar) and hydrophobic (nonpolar) domains for the heterogeneous microenvironment, [$C_8$mim][$PF_6$] ionic liquid was chosen as an example for MD simulation. Simulation was conducted according to the following protocol: Firstly, the ion pairs of [$C_8$mim][$PF_6$] ionic liquid were packed randomly oriented into the simulation box. Eight [$C_8$mim]$^+$ and eight [$PF_6$]$^-$ were packed into the simulation box using the Amorphous Cell Module in Materials Studio Program Package.[6] The size of the simulation boxes were set to 1.5 nm × 1.5 nm × 1.5 nm. After initial setup of the simulation box, the system was first optimized with 1000 steps of the steepest descent minimization. After initial energy minimization, production runs of 5.0 ns duration at constant volume were conducted. The particle mesh Ewald method was used to calculate electrostatic interactions, with a cut-off of 8.5 Å for the separation of the direct and reciprocal space summation.[7,8] The cut-off distance for van der Waals interaction was 8.5 Å, and the parameters of the Lennard-Jones potential for the cross interactions between non-bonded atoms were obtained from the Lorentz-Berthelot combination rule.[9-11] In all MD simulations, the time step of 1.0 fs was used and the coordinates were saved every 1 ps. All atom MD simulations were performed using the Materials Studio Program package.[6] In both energy minimization and MD simulations, the COMPASS force field was used.[12]

Ionic (polar) and hydrophobic (nonpolar) domains in the ionic liquid are obtained from the result of the MD simulation. Supplementary Fig. 6 shows the distribution of polar and nonpolar regions in [$C_8$mim][$PF_6$] ionic liquids, it is found that two domains are formed in ionic liquid, the polar domains are rich in anions and imidazolium rings while the nonpolar domains are rich in alkyl chains, which is in agreement with our previous experimental results.[13,14] The polar domains are formed through electrostatic attractions, whereas the alkyl chains do not participate in the strong interaction network of polar domains, and thus pack together and form the nonpolar domains through van der Waals interactions.

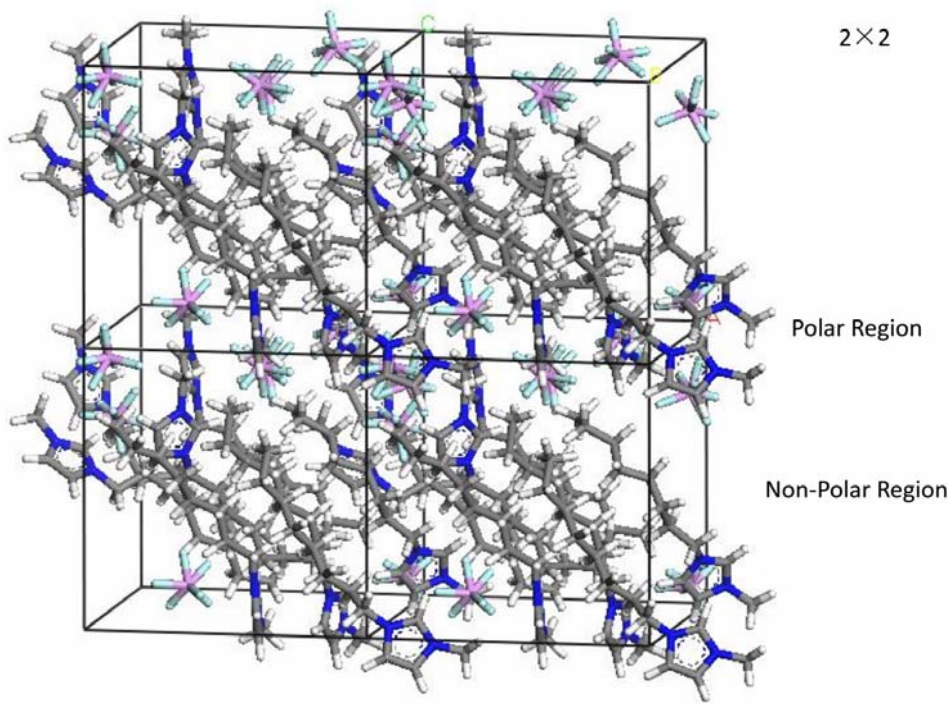

**Supplementary Figure 6.** Distribution of polar and nonpolar domains in ionic liquid [$C_8$mim][$PF_6$] obtained by MD simulation.

### Distribution of $O_2$ in the Ionic Liquid.

Based on the understanding of the ionic (polar) and hydrophobic (nonpolar) domains in the ionic liquid, we build the layer by layer model, where the charged components of imidazolium rings and anions were packed in the ionic domain while the hydrophobic alkyl tails were pointing into the hydrophobic domain (as shown in Supplementary Fig. 6). The effective polarization of oxygen could cause the charge shift in the $O_2$ molecules, and thus enhance the interactions between polarized $O_2$ and the charged components of ionic liquid. However, the polarization of $O_2$ will not take place when oxygen molecules are far away from polar domain in the nonpolar domain unless induced by laser, the polarized $O_2$ in nonpolar domain can then move into polar domain through electrostatic attractions. To reduce the simulation time, we have made some simplification during simulation, for example, $O_2$ molecules were only packed in the polar domain, and internal-built electric field from polar domain is used instead of external laser field since the external light field is 1Hz, ~8ns pulse laser. To properly describe the polarization of oxygen, QM/MM method was used.[15,16] The energy of ionic pair of [$C_8$mim]$^+$ /[$PF_6$]$^-$ and $O_2$ were calculated using B3LYP method with effective core potential.[17,18] Charge distribution in Supplementary Fig. 7 shows the polarization of the $O_2$ at (+0.042, -0.041). The energy of the other parts of the system is calculated using the similar method as described before. Energy and temperature changes in the simulation are shown in Supplementary Figs 8 and 9, respectively.

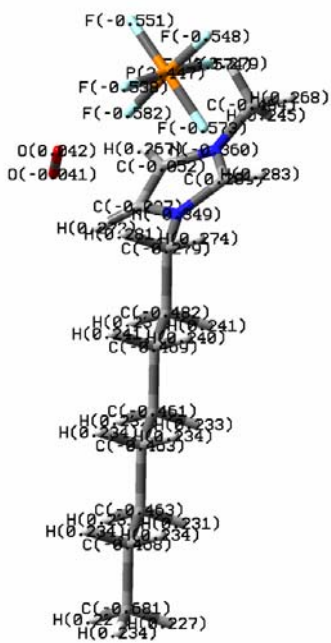

**Supplementary Figure 7.** The charge distributions of atoms are marked in this fig. Polarization of the oxygen is significant. The molecular structures of $[C_8mim]^+/[PF_6]^-$ and $O_2$ are taken from the energy minimized simulation box (shown in Supplementary Fig. 8, 0 ns).

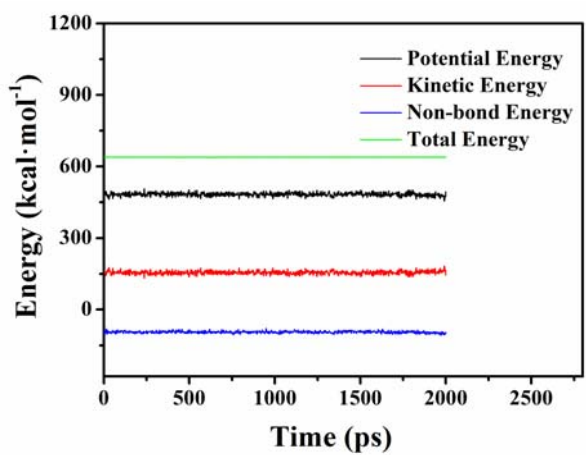

**Supplementary Figure 8.** Energy changes with time in the molecular simulation showing the stable kinetic energy and potential energy within the simulation of 2 ns.

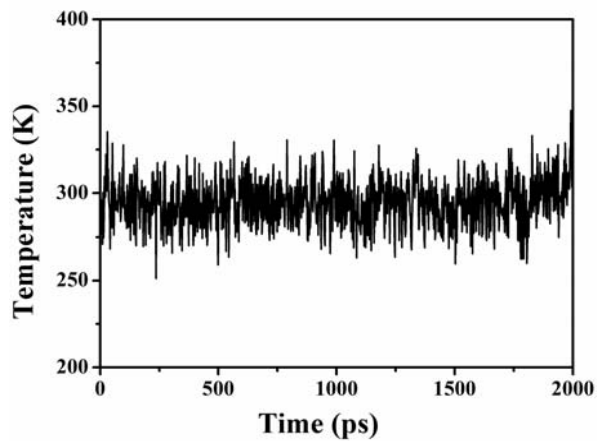

**Supplementary Figure 9.** Temperature change in simulation. During the simulation the temperature remains close around 300 K, which is corresponding to the kinetic energy change in Supplementary Fig. 8.

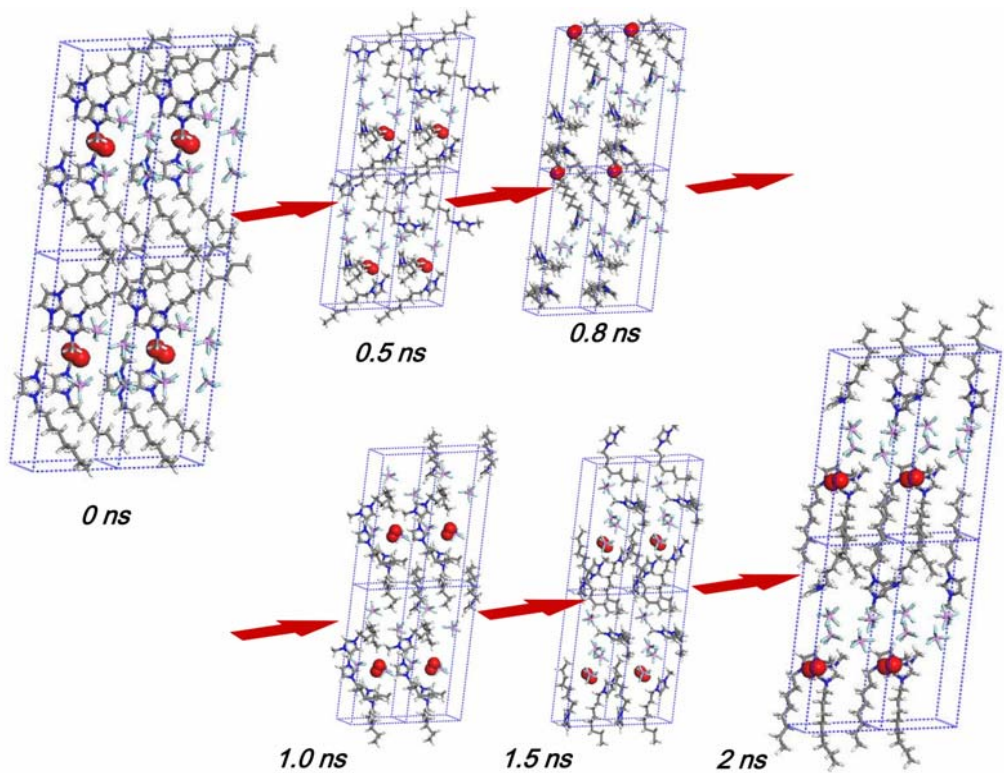

**Supplementary Figure 10.** Structure change in simulation. The energy minimized structure were used as the initial structure of 0.0 ns. To properly describe the polarization of oxygen, QM/MM method was used. The energy of ionic pair of $[C_8mim]^+$ /$[PF_6]^-$ and $O_2$ were calculated using B3LYP method with effective core potential.

Supplementary Fig. 10 shows the evolution of structural changes extracted from MD simulations from 0 to 2 ns. From the simulation, it is found that the volume of the heterogeneous system was reduced. The orientation of $[C_8mim]^+$ changed from slanted to perpendicular, and the $[C_8mim]^+$ molecules were packed more tightly. The position of $O_2$ changed from the very center of the ionic (polar) domain (0.0 ns) to the boundary of the hydrophobic domain (0.8 ns) and finally close to the charged components of the polar domain (2.0 ns), which clearly shows that the polarized oxygen molecules prefer to stay in the polar domain than nonpolar domain and would move close to the charged components in ionic liquid. The evolution process for the oxygen distribution in the heterogeneous microstructure of ionic liquid $[C_8mim][PF_6]$ by MD is further shown in Supplementary Figs 11 and 12. It is found that the polarized $O_2$ molecules are randomly distributed in polar domain of ionic liquids time-by-time.

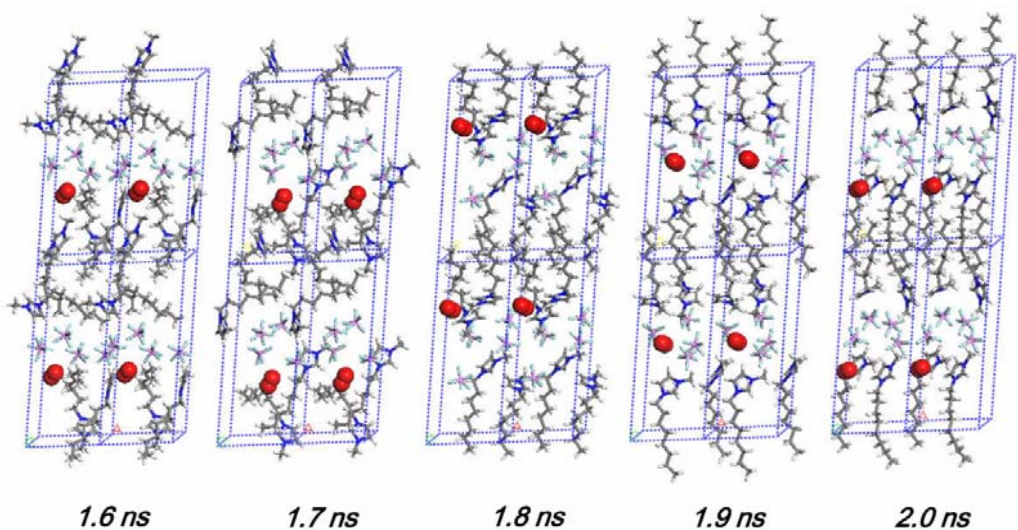

1.6 ns    1.7 ns    1.8 ns    1.9 ns    2.0 ns

**Supplementary Figure 11.** Structure change in simulation. During the simulation, oxygen could stay either in the upper (at 1.8 ns) or in the lower side (at 1.7 and 2.0 ns) of the ionic (polar) domains. The transition states between upper and lower sides of polar domain are also detected at 1.9 ns and 1.78 ns (which is shown in the following fig., Supplementary Fig. 12).

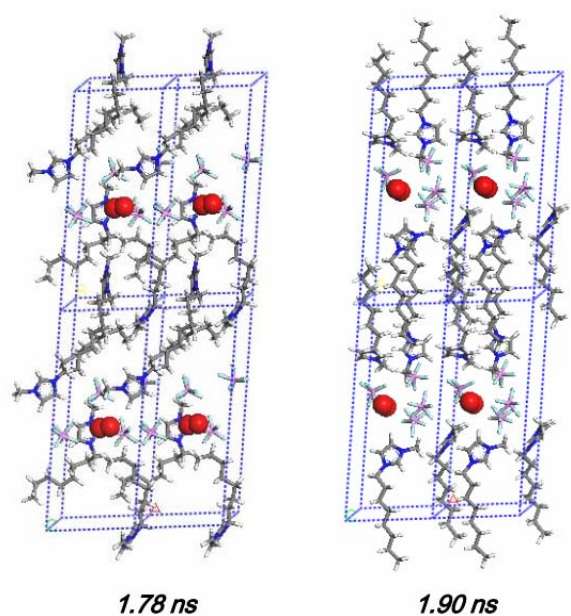

*1.78 ns*    *1.90 ns*

**Supplementary Figure 12.** The transition states when oxygen was moving from the lower side to the upper side of the ionic (polar) domain at 1.78 ns, while at 1.90 ns, it moves back down to the lower side from the upper side.

It is quite significant to see that the oxygen molecules mainly distribute in the polar domains rather than in the nonpolar domains. Even though the interactions between oxygen molecules and the charged components are obviously weaker than the electrostatic attractions between ion pairs, they are much stronger than the ones between alkyl chains and the charged components. This is to say that the polar domains prefer the oxygen from the alkyl chains, and $O_2$ molecules could easily reach to the boundary of polar and nonpolar domains and keep away from the alkyl chains. On the other hand, the interaction between the alkyl chains and the oxygen molecules is even weaker than the interactions within alkyl chains themselves. Furthermore, oxygen molecules could be easily polarized by internal-built ionic field of ionic liquids and move into the polar domain. Therefore, from the point of energy, the process with $O_2$ inserting into the nonpolar domain is quite inferior with external field or internal-built ionic field. In this case, the polarized $O_2$ are randomly distributed in polar domain of ionic liquids.

## Supplementary Note 6

**Rotational Dynamics of C153.**

The reasonable schematic representation of the structural heterogeneities for ionic liquids [C$_n$mim][PF$_6$] ($n$ = 4, 6, 8) investigated by C153 at initially normal conditions is drawn as shown in Supplementary Fig. 13a based on our previous study.[13,14] From Supplementary Fig. 13 and Supplementary Table 5, it is found that two different rotation time constants are obtained by fitting the decays of C153 time-resolved anisotropy in all the ionic liquids [C$_n$mim][PF$_6$] ($n$ = 4, 6, 8) after different amount of laser pulses irradiations. The fast and slow components of C153 rotational dynamics are accordingly attributed to the nonpolar and polar domains with an incompact and compact microstructures in ionic liquids (Supplementary Fig. 13a).[13,14] The C153 rotational behaviors in ionic liquids [C$_n$mim][PF$_6$] ($n$ = 4, 6, 8) are almost unaffected after 10 laser pulses (about 10 seconds) and much similar to the non-irradiation initial cases,[13] and no obvious changes are observed before 100 pulses of laser irradiation. After 300 and 500 pulses, the proportion of the slow component of C153 rotational dynamics in ionic liquid [C$_4$mim][PF$_6$] increases slightly from initial 51% at 10 pulses to 58% at 300 pulses and 57% at 500 pulses, respectively, and that of the fast rotation time constant increases from 0.33 ns to 0.47 and 0.46 ns. In ionic liquid [C$_6$mim][PF$_6$], the rotational behaviors of C153 are almost unaffected even after a longer effect of the laser radiations for 300 and 500 pulses. Regarding the rotational dynamics of C153 in ionic liquid [C$_8$mim][PF$_6$], it is found that after 300 and 500 pulses of laser irradiation, the proportion of slow component increases slightly from initial 51% at 10 pulses to 55% at 300 pulses and further up to 58% at 500 pulses, and the fast rotation time constant increases from 0.47 to 0.53 ns and further up to 0.65 ns.

With a certain free volume, the rearrangement of the microstructure of ionic liquid [C$_4$mim][PF$_6$] with a short side chain has been possibly completed under the local electric field by laser irradiation with about 300 laser pulses, leading to less change of the rotational dynamics of C153 up to 500 pulses irradiation. In the case of ionic liquid [C$_6$mim][PF$_6$], the rotational dynamics of C153 is almost unchanged and the C153 average rotation time is slightly slower even after 300 and 500 pulses of laser irradiations, this is because there is less enough free volume to rearrange the microstructure for the [C$_6$mim][PF$_6$] with the longer hexyl chains compared to ionic liquid [C$_4$mim][PF$_6$] under the external electric field as shown in Supplementary Fig. 13a. For a relative large enough free volume in the heterogeneous structure with the unordered and coiled octyl chains of ionic liquid [C$_8$mim][PF$_6$] (Supplementary Fig. 13a), the ordered pattern could be tightly formed along the electric field direction for 500 pulses of laser irradiation, and the charged components and the ordered interdigitated octyl chains could mainly contribute the ordered arrangement. The average rotation time constant of C153 in ionic liquid $[C_8mim][PF_6]$ increases from 2.84 to 3.06 ns and further increases to 3.25 ns as the laser irradiation increases from 10 to 300 and further to 500 pulses.

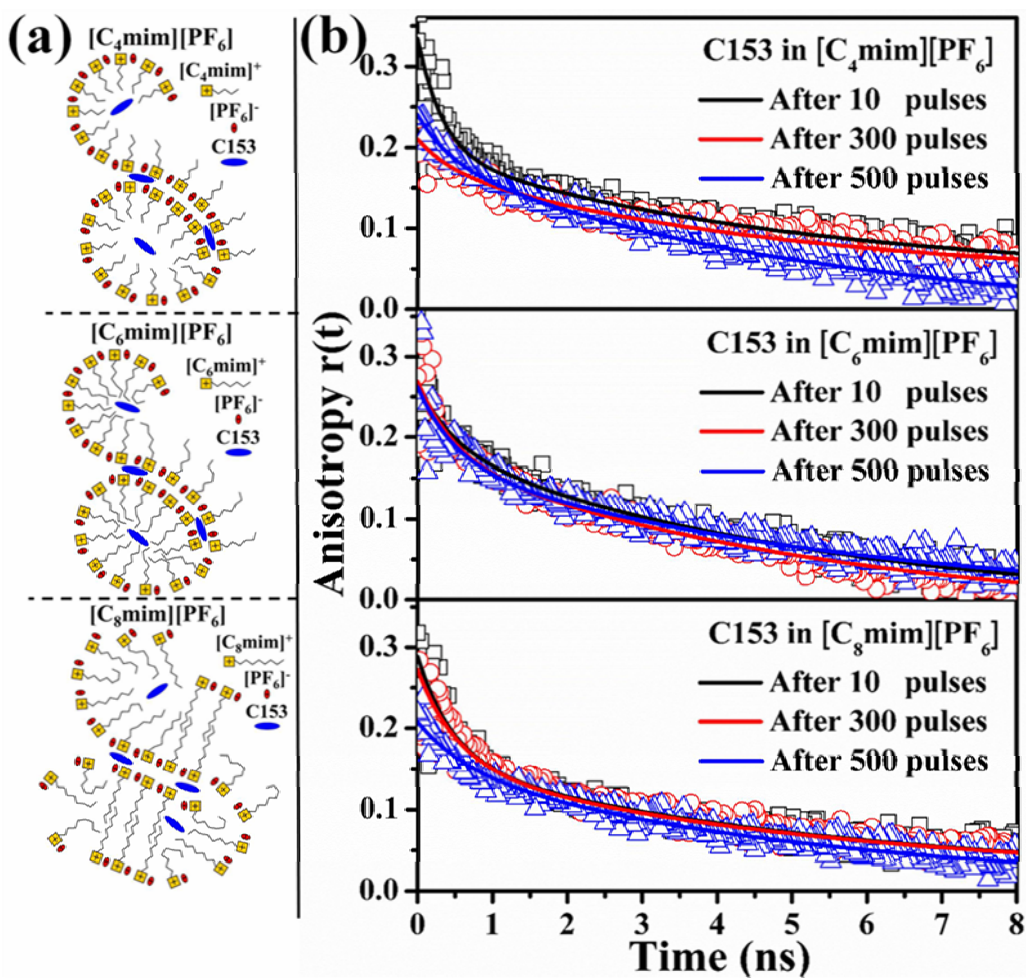

**Supplementary Figure 13.** (a) Schematic representation of C153 in ionic liquids $[C_nmim][PF_6]$ ($n$ = 4, 6, 8) with different heterogeneities at initially normal conditions.[13] (b) Time-resolved fluorescence anisotropy decays $r(t)$ of C153 in ionic liquids $[C_nmim][PF_6]$ ($n$ = 4, 6, 8) after different laser pulse irradiations. The solid lines represent the fitting results.

**Supplementary Table 5.** Rotational relaxation parameters for C153 in ionic liquids $[C_nmim][PF_6]$ ($n$ = 4, 6, 8) after different laser pulse irradiations.

| Ionic Liquids | Laser pulses | Fitting results[a] |
| --- | --- | --- |

| [C$_n$mim][PF$_6$] | for irradiation | $A_1$ | $\tau_1$ (ns) | $A_2$ | $\tau_2$ (ns) | $\tau_r^b$ (ns) |
|---|---|---|---|---|---|---|
| [C$_4$mim][PF$_6$] | After 10 pulses | 0.49 | 0.33 | 0.51 | 4.92 | 2.67 |
| | After 300 pulses | 0.42 | 0.47 | 0.58 | 4.94 | 3.06 |
| | After 500 pulses | 0.43 | 0.46 | 0.57 | 4.93 | 3.01 |
| [C$_6$mim][PF$_6$] | After 10 pulses | 0.45 | 0.48 | 0.55 | 5.28 | 3.12 |
| | After 300 pulses | 0.45 | 0.47 | 0.55 | 5.30 | 3.13 |
| | After 500 pulses | 0.44 | 0.48 | 0.56 | 5.31 | 3.18 |
| [C$_8$mim][PF$_6$] | After 10 pulses | 0.49 | 0.47 | 0.51 | 5.11 | 2.84 |
| | After 300 pulses | 0.45 | 0.53 | 0.55 | 5.13 | 3.06 |
| | After 500 pulses | 0.42 | 0.65 | 0.58 | 5.13 | 3.25 |

[a]The single exponential fitting using $r(t) = a_1 e^{-t/\tau_1}$, The biexponential fitting using $r(t) = \sum_{i=1}^{2} a_i e^{-t/\tau_i}$, [b]The average lifetime from biexponential fitting using $<\tau> = \sum_{i=1}^{2} a_i \tau_i / \sum_{i=1}^{2} a_i$, $A_i = a_i / \sum_{i=1}^{2} a_i$, $\sum_{i=1}^{2} A_i = 1$.

## Supplementary References